\documentstyle[pre,aps,psfig]{revtex}
\def\be{\begin{equation}}
\def\ee{\end{equation}}
\def\bc{\begin{center}}
\def\ec{\end{center}}

\begin{document}
 
\input epsf.sty

\twocolumn[\hsize\textwidth\columnwidth\hsize\csname %
@twocolumnfalse\endcsname

\draft
 
\widetext
\title{Phase transitions in a cluster molecular field approximation}

\author{Hans Behringer$^{1,2}$, Michel Pleimling$^{1}$, and Alfred H\"{u}ller$^{1,2}$}
 
\address{
 $^1$Institut f\"ur Theoretische Physik I, Universit\"at Erlangen-N\"urnberg,
\linebreak D -- 91058 Erlangen, Germany \linebreak
 $^2$Institut Laue-Langevin, F -- 38042 Grenoble, France}
\maketitle
 
\begin{abstract}
Cluster molecular field approximations represent a substantial 
progress over the simple Weiss theory where only one spin is 
considered in the molecular field resulting from all the other 
spins. In this work we discuss a systematic way of improving the
molecular field approximation by inserting spin clusters of variable
sizes into a homogeneously magnetised background. The density of 
states of these spin clusters is then computed exactly. We show
that the true non-classical critical exponents can be extracted 
from spin clusters treated in such a manner. For this purpose a molecular field finite
size scaling theory is discussed and effective critical exponents 
are analysed. Reliable values of critical quantities of various Ising and Potts models
are extracted from very small system sizes. 
\end{abstract}
 
\pacs{05.50.+q, 64.60.-i, 75.10.-b}
 
\phantom{.}
]
 
\narrowtext

\section {Introduction}
The molecular field approximation (MFA) \cite{Wei07} is the method of choice
to gain a first, usually qualitatively correct, understanding of
phase diagrams in cases where exact results are not available.
However,
quantitative predictions derived in the framework of this approximation
compare rather poorly with results obtained by other, more refined,
methods. For example, the molecular field approximation systematically
overestimates the temperatures at which phase transitions take place.
This is of course due to the fact that correlations are completely
neglected in this treatment. Furthermore, the critical exponents,
describing the behaviour of critical quantities in the vicinity
of a critical point, usually do not agree with the molecular
field critical exponents. Above the upper critical 
dimension, however, the molecular field predictions of critical exponents
become exact.

The shortcomings of the original MFA led Bethe \cite{Bet35}
(see also Ref.\ \onlinecite{Pei36} and \onlinecite{Wei48})
to consider a more general treatment. In the Weiss
theory of ferromagnetism, a single spin is considered
in the molecular field which is caused by all the other spins.
This is different in 
the Bethe approach where a small cluster of spins, 
consisting of the central spin and its nearest neighbours, is treated exactly.
The molecular field acts exclusively on the edges of the cluster.
Thus, correlations between a certain number of spins are
included into the calculation.
This is a significant improvement over simple molecular field theory,
yielding for example critical temperatures which are much closer to the correct
values than those obtained from the Weiss theory.

The general idea of inserting a cluster of spins into a homogeneously
magnetised medium led to several applications in the past. M\"{u}ller-Krumbhaar
and Binder, for example, proposed a Monte Carlo method where a molecular field
acts only on the surface spins
of a cluster. This field has to be determined in a self-consistent
way during the course of the simulation \cite{Mul72}.
In a series of papers, Suzuki and coworkers established a general
theoretical framework for extracting non-classical critical exponents
from cluster-mean-field approximations \cite{Suz86a,Suz86b,Suz86c,Suz87,Kat87,Hu87}.
They considered a systematic
series of clusters and showed that the true critical exponents can
be obtained from the amplitudes of mean-field-type singularities.
Various other generalisation schemes
of the one-spin molecular field approximation have been discussed in the 
literature \cite{Gal96,Kam97}.

In this work we present a systematic way of improving
the molecular field approximation by considering spin clusters of
variable sizes embedded into a homogeneously magnetised background.
Hereby, the interactions within the clusters are taken into account
exactly. To this end we compute the density of states as a function of the
internal energy and of the interaction energy with a molecular
field of unit strength. This is a hard task, but when it has been completed 
everything else does no longer demand time-consuming calculations.
The actually applied molecular field itself is determined by imposing a
condition of homogeneity in the central part of the cluster. This approach
is used to evaluate various temperature dependent quantities for 
Ising models defined on different two- and three-dimensional lattices.
A two-dimensional Ising model with nearest and next-nearest neighbour couplings
has also been studied. Results obtained for three-state Potts
models on square and triangular lattices are discussed, too.

With these data we demonstrate that critical quantities (critical 
temperatures as well as critical exponents) 
can be extracted reliably from the cluster molecular field approximation. 
This is already achieved by
inserting very small clusters containing a small number of spins
(the largest cluster studied is formed by 181 spins)
into the molecular field.
Besides analysing effective exponents
we also show that the cluster molecular field data exhibit a remarkable finite size
scaling behaviour which can be exploited for the determination of the true
non-classical values of the critical exponents.

The paper is organised in the following way. In the next Section we present our
method of systematically improving the molecular field approximation. 
We also discuss how to obtain various quantities from this data. In Section III
the molecular field finite size scaling theory is discussed. We thereby use both the reduced
temperature and the reduced energy as possible scaling variables.
Section IV presents our data. We analyse them in different ways and show 
that critical quantities can be obtained
from small clusters in a molecular field.
Finally, in the last Section we 
present our conclusions.

\section{Clusters in a molecular field}
In the following we propose a systematic scheme for improving
the molecular field approximation. This scheme combines the exact
computation of the density of states of small clusters with a
molecular field acting solely on the borders of the clusters.
In Section IV this method is applied to the determination of
non-classical critical quantities of various Ising and three-state
Potts models. Before discussing our approach in detail we will
briefly reiterate the reasoning leading to the molecular 
field self consistency
equation for these two models. 

\subsection{The molecular field approximation}
In zero field the Ising Hamiltonian reads
\begin{equation} \label{Gl:1}
H_I = -J_I \sum_{\left<i,j\right>} \sigma_i\sigma_j
\end{equation}
where $\sigma_i = \pm 1$ is an Ising spin on the lattice. Angular   
brackets denote sums over nearest neighbour pairs and $J_I$ is the   
exchange constant. The
Hamiltonian for one selected spin $i$, given the spin values
$\sigma_{i+\delta}$ of its $z$ neighbours, is
\begin{equation} \label{Gl:2}
\tilde{H}_{I} = - J_I \, \sigma_i \sum_{\delta =1}^{z} \sigma_{i+\delta} .
\end{equation}
If the spin $\sigma_i$ is considered and the rest of the lattice is replaced by a
homogeneously magnetised medium with magnetisation $m$ then the spin variables 
$\sigma_{i+\delta}$ are replaced by the expectation value  $\left< \sigma_{i+\delta} \right> = m$ and
$\tilde{H}_{I}$ becomes $H_{I,MFA}(\sigma_i) = - z J_I \, m \, \sigma_i$ .

At the temperature T the expectation value of $\sigma_i$ in the
molecular field $z J_I \, m$ is:

\begin{equation} \label{Gl:3}
\left<\sigma_i\right> = \frac{Z_+ - Z_-}{Z} = \tanh ( z J_I m /k_BT)
\end{equation}
where $Z_+ =\exp( z J_I m /k_BT)$ , $Z_- =\exp(- z J_I m /k_BT)$
and $Z = Z_+ + Z_-$ is
the partition function for the two states of a single spin.

The condition $\left<\sigma_i\right> = m$ yields the self consistency
equation which can be solved for the temperature

\begin{equation} \label{Gl:4}
k_B T(m)/J_I = \frac{z \, m}{{\mbox {artanh}} (m)} \;.
\end{equation}
Eq. (4) can be inverted or solved graphically if necessary. From the limit
$m \to 0$  of (4) one
finds the MFA critical temperature $k_B T_{c,MFA}/J_I = z$ and for small but
finite $m$ one obtains

\begin{equation} \label{Gl:5}
k_B T(m)/J_I = k_B T_{c,MFA}/J_I -\frac{z}{3} \: m^{1/\beta_{MFA}} + \ldots
\end{equation}
with the critical exponent $\beta_{MFA} = 1/2$.

The derivation of the corresponding self consistency equation
for the three-state Potts model \cite{Wu82} defined by the Hamiltonian
(the spins $\sigma_i$ taking the values 1, 2, 3)
\begin{equation} \label{Gl:6}
H_P = - J_P \sum_{\left<i,j\right>} \delta (\sigma_i, \sigma_j)
\end{equation}
with $\delta (\sigma_i, \sigma_j) = 1$ if $\sigma_i = \sigma_j$
and zero otherwise is somehow more involved, so we only quote
the result \cite{Kih54,Mit74}
\begin{equation} \label{Gl:7}
m = \frac{\exp ( z J_P m /k_BT) - 1}{\exp ( z J_P m /k_BT) + 2}.
\end{equation}
Inspection of the molecular field free energy reveals that at 
the temperature 
\begin{equation} \label{Gl:8}
k_B T_c /J_P = \frac{z}{4 \ln (2)}
\end{equation}
a first order transition takes place where 
the order parameter $m$ jumps from 0 to the value 1/2.
This discontinuous behaviour in the MFA is in contrast to the fact that
the temperature driven phase transition is continuous for the
three-state Potts model in two dimensions.

These highly familiar results have been repeated also for the purpose of
analysing the main defect of the approach. It is the lack of any correlation
between the selected spin $\sigma_i$
and all the other spins in the lattice. This can be partially repaired if a cluster of $N$ spins is inserted
into the homogeneously magnetised medium.
In our approach, as discussed in the following, 
interactions within such clusters are taken into account
exactly whereas the effect of the magnetised background is treated in the
ordinary Weiss approximation. Therefore the correlations
of the central spin with a number of neighbouring spins
are included into the calculation thus leading to greatly improved
results.

\subsection{Cluster molecular field approximation}
In this Section we present our method to calculate physical 
quantities within  an extended molecular field approximation 
based on the idea of Bethe.

Consider a classical spin system on a lattice defined by 
the Hamiltonian 
\begin{equation}
  \label{eq:hamiltonian}
  H = \sum_{\left<i,j\right>} I(\sigma_i,\sigma_j) 
\end{equation}
where $I(\sigma_i, \sigma_j)$ is the interaction energy
of the two neighbouring spins $\sigma_i$ and $\sigma_j$. For the
Ising model this is just the negative product of the spin variables,
i.e. $I_I(\sigma_i, \sigma_j) = -J_I\sigma_i \sigma_j$, for
the $q$-state Potts model ($\sigma_i = 1, \ldots , q$) the interaction
energy is given by $I_P(\sigma_i, \sigma_j) =
-J_P\delta(\sigma_i, \sigma_j)$. 
Suppose one has constructed a cluster of size $N$ by some method (see below) 
which is embedded in the original lattice.     
For this cluster the Hamiltonian 
(\ref{eq:hamiltonian}) is replaced by the molecular field Hamiltonian
\begin{equation}
  \label{eq:hamil_mf}
  H_N = \sum_{\left<i,j\right>}I(\sigma_i, \sigma_j)
  +\mu\sum_{\{i, k\}}I(\sigma_i, \sigma_k) \delta (\sigma_k, 1) 
\end{equation}
where
$\left<i,j\right>$ indicates a summation over 
nearest neighbour pairs within the cluster. 
The second term in (\ref{eq:hamil_mf}) represents the interaction of the cluster 
with the homogeneously magnetised background. The sum runs over all pairs of 
nearest neighbour spins where the spin $\sigma_i$ is from inside the cluster and its 
partner $\sigma_k$ is from outside. This is denoted by the curly brackets. For 
convenience the $\sigma_k$ are set equal to 1 and their common expectation value 
$\left< \sigma_k\right>=\mu$ is placed outside the summation. 
%
%
%The second sum in (\ref{eq:hamil_mf}) denoted by curly brackets 
%runs over all nearest neighbour 
%$pairs of border spins $\sigma_i$ of the cluster with partners $\sigma_k$ 
%outside.   
%These spins interact with the homogeneous 
%background field $\mu$ which is associated with a fixed spin 
%state $\sigma_k = 1$ of the sites 
%outside the cluster. 
%This contribution represents the interaction 
%with the homogeneous background magnetisation $\left< \sigma_k\right>=\mu$. 
%Note that the cluster is embedded in 
%the original lattice so that the spins at the corners 
%interact with the field $\mu$ through three sites whereas the 
%spins on the edges interact only with two sites outside the cluster 
%in case of a square lattice. 
The Hamiltonian (\ref{eq:hamil_mf}) of the extended molecular field 
approximation treats the statistical mechanics of the spin system 
within the cluster exactly and the interaction of the cluster 
with the background is modelled by a molecular field that 
has to be determined in a self consistent way.
In order to calculate physical quantities in the cluster molecular 
field approximation it is advantageous to work out the density   
of states 
\begin{equation}
  \label{eq:dos}
  \Omega_{\sigma_0, \sigma_1}(E,B)=\sum_{\Gamma_N(\sigma_0,\sigma_1)}
  \delta(H_C,E)
  \delta(H_B, B) \;,
\end{equation}
with the internal energy of the cluster 
\begin{equation}
  H_C = \sum_{\left<i,j\right>}I(\sigma_i, \sigma_j)
\end{equation}
and the interaction energy 
\begin{equation}
  H_B = \sum_{\{i, k\}}I(\sigma_i, \sigma_k) \delta (\sigma_k, 1)
\end{equation}
of the cluster with the spins outside. 
The summation in (\ref{eq:dos}) runs over all $\Gamma_N(\sigma_0,\sigma_1)$ states of the
cluster with fixed spin value $\sigma_0$ of the central spin and
$\sigma_1$ of one of its $z$ equivalent neighbours. 
For classical discrete spin models the density of states 
is just the number of configurations with specified values $E$, $B$, 
$\sigma_0$ and $\sigma_1$.  

To work out physical quantities within this approximation scheme the density 
of states is evaluated exactly so that these quantities can be computed with 
arbitrary accuracy. The technical details of computing the density of states 
(\ref{eq:dos}) for a given cluster size $N$ are summarised in the 
appendix. The size of the clusters which can be handled exactly is limited by 
the available computational capacities. 
To treat bigger clusters a recently proposed, highly efficient 
Monte Carlo algorithm to compute the density of states could be used \cite{Hueller02}. However, a 
Monte Carlo calculation has to be performed with great care to achieve the desired  
accuracy.
 
\begin{figure}[t]
\centerline{\psfig{figure=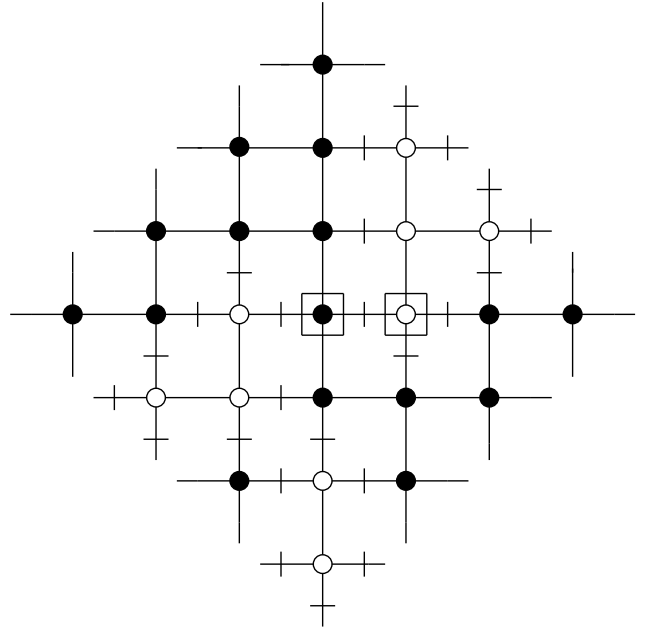,angle=270,width=3.3in}}
\caption{One of the $2^{25}$ configurations of an Ising cluster of 25 spins.
Spin up $(\sigma_i = +1)$ is denoted by a full circle, spin down
$(\sigma_i = -1)$ by an open circle. Broken bonds are marked with a
little bar. Here $E = -6$ and $B = -10$. The central spin $\sigma_0
=+1$ and one of its first neighbours $\sigma_1 = -1$ are identified by
little square. This configuration contributes one count to $\Omega
_{+-}(-6,-10) = 57642$.}
\end{figure}

Clusters of increasing particle numbers are selected in two different ways. 
A possible systematic series of clusters
is constructed by choosing a spin to be the central spin and successively
including shells of nearest neighbours. Hence, the first cluster of $z+1$
spins is made up  of the central spin and its $z$ nearest
neighbours.
The next cluster consists of the first cluster and
the shell of nearest neighbours of the border spins of the first cluster.
By this way a new cluster is made up of the previous one
together with the nearest neighbours of the boundary sites
of the original cluster. 
This method has the advantage that assemblies of the same shape are build up.
Fig. 1 displays the
cluster of 25 spins emerging from this method for a two-dimensional
square lattice.
For the topology of the square lattice a cluster consists of $L$ rows of length $L$
perpendicular to the bisector of a quadrant of the plane and
of $L-1$ short rows of length $L-1$ between the longer ones. Hence the
number of spins in a cluster is given by $N=L^2+(L-1)^2$. The construction
of successive clusters leads to a series of clusters with 5, 13, 25,
41, 61, 85, 113, 145, and 181 spins for the two-dimensional square lattice 
which can be handled with the available computational capacities.
For the triangular lattice clusters of 7, 19, 37, 61, 91, and 127 spins emerge.
In this geometry the basic row and the central row have the lengths $L$ and
$2L-1$, respectively.
However to avoid additional computational effort due to the length of the
central row for the treatment of 
the triangular lattice we alternatively use the clusters 
of the square lattice by inserting additional bonds between the sites of 
the rows of lengths $L$ and $L-1$ so that all sites have the coordination
number 6 of the triangular topology.  
In three dimensions the equivalent procedure would be restricted to only one, 
or at most two system sizes by the available storage capacities. This imposes 
a less systematic approach. The clusters are constructed by adding successive 
shells of neighbours of the central spin. With the nearest neighbours 
of the central spin one obtains 
a cluster of 7 spins, with the next nearest neighbours the size is 19, then 
27, 33, 57, and so on. The cluster of 57 spins that consists of the central 
spin and includes all spins up to the 5th neighbours was the largest one 
which could be handled in three dimensions. This way of selecting clusters for the 
three-dimensional case has the disadvantage that the 
various assemblies do not have the same geometry.  
 
Once the density of states is known physical quantities 
are easily evaluated. For example the expectation value of the 
central spin is given by 
\begin{equation}
  \label{eq:extval_s0}
  \left<\sigma_0\right> = \sum_{\sigma_0,\sigma_1} 
  \sigma_0  
  Z_{\sigma_0, \sigma_1} / Z
\end{equation}
with the restricted partition functions 
\begin{equation}
  Z_{\sigma_0, \sigma_1} = \sum_{E,B}\Omega_{\sigma_0,\sigma_1}
  (E, B) \exp\left(-(E+\mu B)/k_BT \right)
\end{equation}
and the total partition function
\begin{equation}
  Z = \sum_{\sigma_0,\sigma_1} Z_{\sigma_0, \sigma_1} \;.
\end{equation}
Similarly the expectation value of the spin $\sigma_1$ 
is computed from (\ref{eq:dos}). 
The unknown homogeneous magnetisation $\mu$ of the sites outside the cluster 
must be determined self consistently. When the expectation value 
$\left<\sigma_0\right>$ of the central spin is calculated 
as a function of $\mu$ and $T$ the temperature might be adjusted 
such that $\left<\sigma_0\right>$ is equal to 
$\mu$: 
\begin{equation}
  \label{eq:selbstkon}
\left<\sigma_0\right>(T) = \mu (T) \; . 
\end{equation}
Thereby the rigid external magnetisation is moved further away from the central spin and 
an improvement of the plain molecular field approximation is observed \cite{Kat87}. 
This is the usual condition to 
determine the molecular field discussed in the literature. 
Nevertheless only the central spin is  
directly related to the rigid background magnetisation through the 
self-consistency equation so that the expectation values of the central spin 
and one of its neighbour spins are different. In our approach the condition 
\begin{equation}
  \label{eq:selbstkonsist}
  \left<\sigma_0\right>(T) = \left<\sigma_1\right>(T) 
\end{equation}
 is used instead. This condition, which implicitly determines $\mu$ as a
function of $T$,  ensures that the spin in the centre of the 
cluster and its neighbouring spins have the same expectation value so 
that at least in the central part of the cluster the equivalence of the 
spins is implemented. 
%The cluster molecular field scheme with a self 
%consistency condition like (\ref{eq:selbstkonsist}) has 
%an additional advantage. 
The spins in the centre of the cluster are not directly 
coupled to the rigid background magnetisation. In the 
extended molecular field approximation based on 
condition (\ref{eq:selbstkonsist}), fluctuations 
of the expectation value of  a spin near the border of the cluster  
are suppressed due to the direct coupling to the homogeneous 
background magnetisation whereas the central spin can develop 
comparatively large fluctuations. The behaviour of the central spin 
in the approximation used is consequently most closely related
to the corresponding behaviour of a spin in the infinite system.
The same holds for the interaction energies of bonds in the centre  
compared to those of bonds near the boundary of 
the cluster. 
%The rather weak coupling to the molecular field through the 
%interaction with the other spins of the cluster has the consequence that 
%large fluctuations can emerge within this cluster improving the 
%results which can be obtained from a series of successively increasing clusters. 
These aspects suggest that the calculation of physical quantities should 
be restricted to the central part of the cluster where the condition of homogeneity is 
rather well fulfilled. 

\begin{figure}
\centerline{\psfig{figure=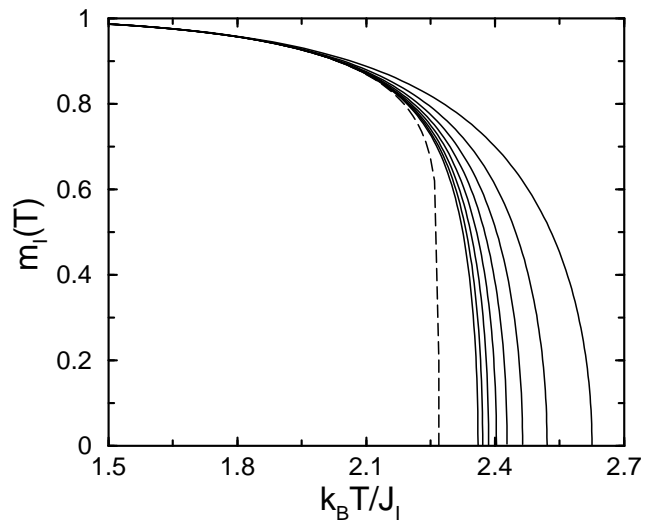,width=3.3in}}
\caption{Order parameters as function of temperature (full lines) as obtained
from the cluster molecular field approximation for various two-dimensional
Ising models defined on a square lattice. System sizes from right to left: 13,
25, 41, 61, 85, 114 145, and 181 spins. The spontaneous magnetisation of the
infinite system is shown as a dashed line. For systems treated
in the cluster molecular field approximation the order parameter sets in
abruptly at a pseudo-critical temperature, see Table I. At low temperatures
the finite size order parameters follow closely the spontaneous magnetisation of the
infinite system.}
\end{figure}

After solving the implicit equation (\ref{eq:selbstkonsist}) for the 
molecular field all physical quantities can be evaluated using the density 
of states (\ref{eq:dos}). The order parameter of the system is determined 
from $\left<\sigma_0 \right>$. Note that in the limit $T \to 0$ the interaction of the 
border spins of the cluster with the homogeneously magnetised background 
forces all spin variables to be in the spin state $\sigma = 1$ of the spins
outside the cluster. For the 
Ising model the order parameter is given by 
\begin{equation}
  \label{eq:ising_ord}
  m_I(T) = \left< \sigma_0 \right> (T) 
\end{equation}
and for the $q$-state Potts model one has 
\begin{equation}
  \label{eq:potts_ord}
  m_P(T) = \frac{q\left< \sigma_0 \right> (T)-1}{q-1} \;.
\end{equation}
%This definition which restricts the order parameter to the properties of the central 
%part of the finite cluster is motivated by the above discussion of 
%the development of comparably large fluctuations in the centre.   

In the cluster molecular field approach the order parameter is 
exactly zero at high temperatures. When decreasing the 
temperature the order parameter of the Ising model exhibits a sudden onset at the temperature $T_{c, N}$ 
indicating a transition to a magnetised phase. The abrupt appearance of a finite 
order parameter defines the pseudo-critical temperature of the finite system.     
Fig. 2 illustrates this temperature dependent behaviour of the order parameter
in the case of the square lattice Ising model. It is instructive to compare
the pseudo-critical temperatures resulting from equations 
(\ref{eq:selbstkon}) and (\ref{eq:selbstkonsist}). For a cluster
of $N=145$ spins defined on the square lattice our approach (\ref{eq:selbstkonsist})
leads  to the value $k_B T_{c,N}/J_I = 2.37096$ ($T_{c,N}/T_c=1.045$, where $k_B T_c/J_I =
2/\ln(1+\sqrt{2})$ is the critical temperature of the two-dimensional Ising model
defined on the square lattice),
which should be compared to the value 2.57524 ($T_{c,N}/T_c= 1.135$) obtained when using the
self-consistency equation (\ref{eq:selbstkon}) \cite{Kat87}. A similar remarkable improvement
is also achieved for three-dimensional clusters: for $N=57$ spins our pseudo-critical 
temperature is $k_B T_{c,N}/J_I = 4.6627$ ($T_{c,N}/T_c=1.034$), 
much closer to the value $4.5115$ of the infinite system than
the value $5.0018$ ($T_{c,N}/T_c=1.109$) resulting from Eq. (\ref{eq:selbstkon}) \cite{Kat87}. 
The pseudo-critical 
temperatures of the Ising model for various lattice topologies are summarised in Table I and Table 
II for the two-dimensional and the three-dimensional systems, respectively. 

In Fig. 3 we show the order parameter for the three-state Potts model defined on a square lattice 
as a function of temperature for a series of cluster sizes. For all 
clusters a jump in the order parameter is observed, thus disclosing
the discontinuous character of the phase transition in the MFA.
This jump gets smaller when the number of cluster spins is increased reflecting
the softening of the first order character of the phase transition with 
increasing cluster size. In the limit $N \longrightarrow \infty$ the jump will disappear
completely in accordance with the continuous phase transition taking place in the 
bulk three-state Potts model. The temperature at which the jump occurs is deduced from the 
behaviour of the free energy. The values of these transition temperatures are displayed in Table III 
for the square and the triangular lattice.

We estimate the internal energy from  
the bonds of the central spin with its $z$ neighbour spins which yields
\begin{equation}
  \label{eq:int_ener}
  u(T) = \frac{z}{2}\sum_{\sigma_0, \sigma_1}
  I(\sigma_0, \sigma_1) 
  Z_{\sigma_0, \sigma_1}/Z \;.
\end{equation}

\begin{table}
\caption{The pseudo-critical temperatures $T_{c, N}$ of the two-dimensional Ising model defined on the
square (sq) and the triangular (tr) lattice.}
\begin{tabular}{|c|c|c|}
$N$ & sq & tr \\
\hline\hline
$\infty$ & $2.26919$ & $3.64096$  \\
 & & \\
\hline
$181$ & $2.36008$ & $\ldots$\\
\hline
$145$ & $2.37096$ & $3.82574$\\
\hline
$113$ & $2.38480$ & $3.85042$\\
\hline
$85$ & $2.40294$ & $3.88338$\\
\hline
$61$ & $2.42778$ & $3.92602$\\
\hline
$41$ & $2.46380$ & $3.99458$\\
\hline
$25$ & $2.52078$ & $4.0804$\\
\hline
$13$ & $2.62534$ & $4.27958$\\
\hline
$5$ & $2.88539$ & $4.72088$\\
%\hline
%$1$ & $4.00000$ & $6.00000$\\
\end{tabular}
\end{table}
\vspace*{1cm}

\begin{table}
\caption{The pseudo-critical temperatures $T_{c, N}$ of the three-dimensional Ising model
on the simple cubic (sc) lattice.}
\begin{tabular}{|c|c|}
$N$ & sc \\
\hline\hline
$\infty$ & $4.5115$  \\
 & \\
\hline
$57$ & $4.66266$ \\
\hline
$33$ & $4.69361$ \\
\hline
$27$ & $4.69367$ \\
\hline
$19$ & $4.74872$ \\
\hline
$7$ & $4.93261$\\
\end{tabular}
\end{table}
\vspace*{1cm}

\begin{table}
\caption{The transition temperatures of the three-state Potts models defined on the
square (sq) and the triangular (tr) lattice.}
\begin{center}
\begin{tabular}{|l|l|l|}
$N$ & sq & tr \\
\hline\hline
$\infty$ & $0.99497$ & $1.58493$  \\
 & & \\
\hline
$85$ & $1.01937$ & $1.62849$\\
\hline
$61$ & $1.02396$ & $1.63619$\\
\hline
$41$ & $1.03051$ & $1.64813$\\
\hline
$25$ & $1.04074$ & $1.66267$\\
\hline
$13$ & $1.05957$ & $1.69818$\\
\end{tabular}
\end{center}
\end{table}

\begin{figure}[t]
\centerline{\psfig{figure=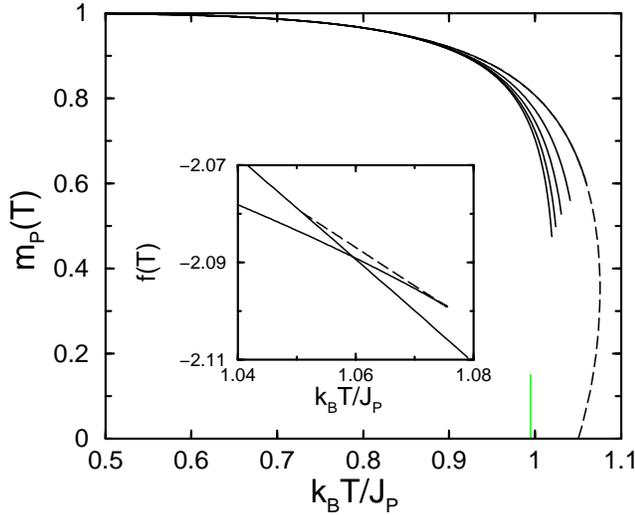,width=3.3in}}
\caption{Temperature dependent order parameters of three-state square lattice Potts models
derived within the cluster molecular field approximation (full lines). The exactly treated clusters
contain (from right to left) 13, 25, 41, 61, and 85 spins. The discontinuous character of
the phase transition
in the cluster molecular field approximation is displayed by the jump of the order parameter.
The unstable solution for a system with 13 spins is shown as a dashed line. The grey line
indicates the location of the critical point of the infinite system: $k_BT_c/J_P
= 1/\ln(\sqrt{3}+1) \approx 0.995$.
Inset: free energies obtained for a cluster with 13 spins within the cluster molecular
field approximation. The free energy of the stable high and low temperature phases are
shown as full lines whereas the dashed line is the free energy of the unstable solution.
The intersection of the full lines yields the transition temperature listed in Table III.}
\end{figure}

\noindent
From the internal energy one now gets the specific heat 
\begin{equation}
  \label{eq:spez}
  c(T) = \frac{du(T)}{dT} \;.
\end{equation}
The specific heat is also related to the entropy of the central spin through 
the relation 
\begin{equation}
  \label{eq:spez_entr}
  c(T) = T\frac{ds(T)}{dT} \;.
\end{equation}
This allows the computation of the free energy 
\begin{equation}
  f(T) = u(T) - Ts(T)
\end{equation}
from the knowledge of the internal energy. Solving equation (\ref{eq:spez_entr}) for the 
entropy $s(T)$ one obtains the free energy 
\begin{equation}
  \label{eq:frei_ener}
  f(T) = u(T) - T \int_{T_0}^T d\tau \frac{c(\tau)}{\tau}
\end{equation}
with $T_0$ being some arbitrarily chosen temperature representing the constant of integration.
The free energy of the Potts model for the cluster with 13 spins in the vicinity of the 
transition point is shown in the inset of Fig.\ 3. The solution of 
(\ref{eq:potts_ord}) that minimises the free energy is physically stable. This leads to the kink of the 
free energy at the transition temperature again indicating the discontinuity of the phase transition 
for finite $N$.

Next we consider the response of the central spin to an external field that acts 
exclusively on the central spin itself.
In order to evaluate this zero field susceptibility one has to differentiate the magnetisation with 
respect to a magnetic field that acts only on the central spin variable $\sigma_0$:  
\begin{equation}
  \label{eq:suszept}
  \chi(T) = \lim_{h \to 0} \frac{\partial m(T, h)}{\partial h} \;.
\end{equation}
To obtain an analytic expression for the susceptibility of the Ising model we 
add a magnetic field term to the Hamiltonian:
\begin{equation}
  H_I \rightarrow H_I - h\sigma_0 \;.
\end{equation}
When evaluating the susceptibility one has to bear in mind that 
one gets a field-dependent molecular field $\mu=\mu(T,h)$ due to this 
additional external field $h$. This has 
the consequence that (\ref{eq:suszept}) depends on the derivative of $\mu$ 
with respect to the magnetic field $h$. However, the order parameter $m$ is 
given by the two equivalent equations $m = \left<\sigma_0\right>$ and 
$m = \left<\sigma_1\right>$ so that one of them can be used to eliminate 
$\partial \mu / \partial h$ from (\ref{eq:suszept}). Differentiating 
$\left < \sigma_0 \right>$ with respect to $h$ one obtains the expression
\begin{equation}
  \chi(T, h) = \frac{1}{k_BT} \left[ \left<\sigma_0^2\right>-\left<\sigma_0\right>^2+\frac{\partial \mu}{\partial h}
      (\left<\sigma_0\right>\left<B\right>-\left<\sigma_0 B\right>) \right] \;.
\end{equation}
On the other hand differentiation of $\left < \sigma_1 \right>$ yields
\begin{equation}
  \chi(T, h) = \frac{1}{k_BT} \left[ \left<\sigma_0\sigma_1\right> -\left<\sigma_0\right>^2+\frac{\partial \mu}{\partial h}
      (\left<\sigma_0\right>\left<B\right>-\left<\sigma_1 B\right>) \right] \;.
\end{equation}
Combining these two relations in order to eliminate the derivative of the molecular 
field results in the desired expression of the zero-field susceptibility 
for temperatures below $T_{c,N}$:
%\begin{equation}
%  \label{eq:sus}
%  \chi(T) = \beta\left[\frac{m\left<B\right>\left (\left<\sigma_0^2\right>-\left<\sigma_0\sigma_1\right> \right)- \left<\sigma_1B\right>
%    + \left<\sigma_0\sigma_1\right>\left<\sigma_0B\right>}{\left<\sigma_0B\right>-\left<\sigma_1B\right>}
%  -m^2 \right]\;.
%\end{equation}
\begin{eqnarray}
  \label{eq:sus_ising}
  \chi(T)& =& \frac{1}{k_BT}\left[\left<\sigma_0^2\right> -
  \left<\sigma_0\right>^2
   \right. \nonumber \\
    &&- \left.\frac{(\left<\sigma_0\right>\left<B\right>-\left<\sigma_0B\right>)
      (\left<\sigma_0^2\right>- \left<\sigma_0\sigma_1\right>)}
    {\left<\sigma_0B\right>-\left<\sigma_1B\right>}\right] \;.
\end{eqnarray}
The first two terms in the square brackets represent the ordinary fluctuation 
formula for the susceptibility. The additional third term arises from the
coupling to the molecular field. 
Its denominator vanishes in the limit $T \to T_{c, N}$ 
yielding a diverging susceptibility for $T \to T_{c,N}$ as expected for the 
molecular field approach. Fig.\ 4 displays the zero field susceptibility of the 
two-dimensional Ising model defined on the square lattice for different cluster sizes.

\begin{figure}[t]
\centerline{\psfig{figure=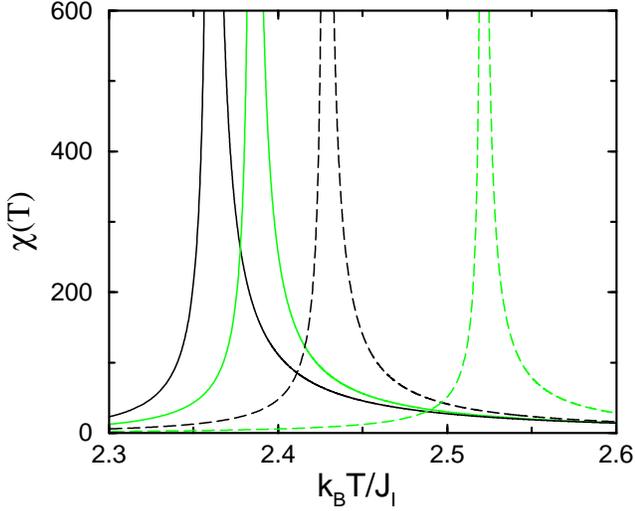,width=3.3in}}
\caption{Zero field susceptibilities as function of temperature obtained from the
cluster molecular field treatment of two-dimensional square lattice Ising clusters,
see Eq.\ (\ref{eq:sus_ising}). The susceptibilities diverge at the pseudo-critical
temperatures. The Ising clusters contain 25, 61, 113, and 181 spins.}
\end{figure}
 
For the $q$-state Potts model the magnetic field acts on the central spin 
only if it is in the spin state $\sigma$ of the sites outside the cluster:
\begin{equation}
  H_P \rightarrow H_P - h \delta(\sigma_0, \sigma) \;.
\end{equation}
The susceptibility of the Potts model for temperatures below the 
pseudo-critical point of the finite cluster is then given by
\begin{eqnarray}
  \label{eq:sus_potts}
  \chi(T)& =& \frac{q}{q-1}\frac{1}{k_BT}\left[\left<\sigma_0 \delta(\sigma_0, \sigma)\right> 
    - \left<\sigma_0\right>\left<\delta(\sigma_0, \sigma)\right>
    \right. \nonumber \\
    && \hspace*{-1cm} - \left. \frac{(\left<\sigma_0\right>\left<B\right>-\left<\sigma_0B\right>)
      (\left<\sigma_0\delta(\sigma_0, \sigma)\right>- 
      \left<\sigma_1\delta(\sigma_0, \sigma)\right>)}
    {\left<\sigma_0B\right>-\left<\sigma_1B\right>}\right] \;.
\end{eqnarray}
 This expression for the Potts model has the same structure as the
corresponding expression (\ref{eq:sus_ising}) of the Ising model. 
However the susceptibility  of the Potts model does not diverge 
as the molecular field displays a jump at the transition temperature.

\section{Molecular field finite size scaling theory}
In this Section we present a phenomenological finite size scaling theory for
our cluster molecular field approximation. We consider both the reduced temperature 
and the reduced energy as scaling variables and discuss the relative size of the 
corrections to scaling. Our data are then analysed in the following Section 
using the framework of the molecular field finite size scaling to be exposed
in the following subsection.  

\subsection{Finite size scaling for the temperature picture}
The spontaneous magnetisation of the cluster molecular field approximation 
displays the characteristics of spontaneous symmetry breaking.
This allows an 
unambiguous definition of the pseudo-critical temperature $T_{c, L}$ 
of the finite system of linear extension $L$ taken to be $N^{1/d}$ where $d$
is the dimensionality of the lattice. 
The spontaneous 
magnetisation near this pseudo-critical temperature then behaves like 
\begin{equation}
  \label{eq:finite_mspont}
        m_{sp, L}(t') = A_{L}(-t')^{\beta'}
\end{equation}
where the reduced temperature $t'$ is defined as
\begin{equation}
  \label{eq:redtemp}
        t' = \frac{T - T_{c, L}}{T_{c, L}}
\end{equation}
and where $\beta' = 0.5$ is the molecular field critical exponent. 
For increasing clusters the influence of the molecular field 
which acts only on the boundary becomes weaker so that 
the approximation eventually becomes exact in the limit $L \to \infty$.    
In the infinite system the spontaneous magnetisation vanishes 
like 
\begin{equation}
  \label{eq:sponmag}
        m_{sp}(t) = A(-t)^{\beta}
\end{equation}
for small reduced temperatures $t$ which is related to the critical temperature 
$T_c= T_{c, \infty}$ of the infinite system by
\begin{equation}
  t = \frac{T-T_c}{T_c} \;.
\end{equation}     
The fractional 
shift $\tau_{L}$ is given by \cite{Barber83} 
\begin{equation}
  \label{eq:fracshift}
        \tau_{L} = \frac{T_{c}-T_{c, L}}{T_{c}} 
        \sim -DL^{-\lambda}
\end{equation}      
with the shift exponent $\lambda = 1/\nu$ where $\nu$ is the critical
exponent of the correlation length. Equation (\ref{eq:fracshift}) 
and (\ref{eq:redtemp}) are combined to yield 
\begin{equation}
  \label{eq:tzuttilde}
        t' = t + \tau_{L} + O(\tau^2_L)
\end{equation}
for large $L$.
Equation (\ref{eq:fracshift}) can be used to obtain an approximation of the transition temperature 
in the thermodynamic limit. This is exploited in Section IV.

For sufficiently small $t'$ we assume that the spontaneous 
magnetisation of the finite system of extension $L$ is given by 
\begin{equation}
  \label{eq:scaling}
        m_{sp, L}(t') = L^p \phi(t' L^{1/\nu}) \;.
\end{equation}
The function $\phi(x)$ is universal showing no further $L$-dependence.  
The exponent $p$ in 
(\ref{eq:scaling}) is obtained by taking the limit $L \to  \infty$ for fixed temperature 
$T \neq T_c$. In this limit equation (\ref{eq:scaling}) reduces to (\ref{eq:sponmag}) yielding 
\begin{equation} 
        \phi(x) \sim x^{\beta} \mbox{ as } x \to \infty \;.
\end{equation}
The residual $L$ dependence is cancelled out if $p=-\beta/\nu$.
Hence the finite size scaling relation 
\begin{equation}
	\label{eq:scaling_mag}
         L^{\beta/\nu} m_{sp, L}(t') = \phi(t' L^{1/\nu})
\end{equation}
follows. 

At this point we have to mention that an unambiguous
definition of the linear extension of our finite clusters is not possible.
This crucial fact was already pointed out 
in \cite{Suz86b}. 
For our investigations we choose $L = N^{1/d}$ but other definitions are also possible. For the 
three-dimensional systems our choice is especially problematic as the clusters do 
not have the same geometry.     
We will come back to this point in Section IV when discussing our numerical data.
%In Section IV the appearance of this effective exponent $\nu^{\prime}$ will be further discussed for both the 
%two-dimensional Ising systems with identical cluster geometries and the three-dimensional Ising system with 
%changing geometries. 

From equation (\ref{eq:finite_mspont}) follows that the universal function $\phi$ behaves like 
a square root for small scaling variables, i.e. 
\begin{equation}
  \phi(t' L^{1/\nu}) = A_{\phi}\sqrt{-t' L^{1/\nu}} \;,
\end{equation}
for $t' \to 0$ and fixed $L$.
Taking the limit $t' \to 0$ and comparing 
with relation (\ref{eq:scaling}) the finite size scaling law
\begin{equation}
	\label{eq:suzuki_scal}
  A_L = A_{\phi}L^{\frac{\beta'-\beta}{\nu}}
\end{equation}
for the amplitudes of the magnetisation of the finite systems is derived. 

In an analogous way we obtain a finite size scaling relation for the susceptibility. In the molecular 
field approach the susceptibility diverges with the molecular field critical exponent
$\gamma' = 1$. 
We then have the scaling relation 
\begin{equation}
   \label{eq:scaling_sus}     
	L^{-\gamma/\nu}\chi_L(t') = \psi(t' L^{1/\nu})  \;,
\end{equation}
where for small scaling variables the universal function $\psi$ is given by 
\begin{equation}
	\psi(t' L^{1/\nu}) = A_{\psi}\,(t' L^{1/\nu})^{-1} \;.	
\end{equation}
One also derives, in complete analogy with Eq.\ (\ref{eq:suzuki_scal}), a
finite size scaling law involving the finite size amplitudes of the susceptibility.
Note that this relation was already used in references \cite{Suz86b,Kat87} to get estimates 
for the critical exponent $\gamma$. All these relations show that the critical exponents of 
the infinite system are encoded in the amplitudes of the corresponding quantity of the 
finite system. It is precisely this fact which will allow the determination of critical 
exponents from the cluster molecular field approximation.
It is also worth noting that in the present approach
the critical temperature of the infinite system is not needed
for the determination of critical exponents.
%that the critical exponents 
%observed within this approximation differ from the true exponents of the model. Furthermore one 
%does not need the critical temperature of the infinite system to extract critical exponents 
%from relation (\ref{eq:scaling_mag}). 
%
%In the following Section finite size scaling relations are used 
%to extract estimates for the true critical exponents from the finite systems treated in the 
%extended molecular field approximation. It will be seen that the molecular 
%field finite size scaling relations allow a determination of critical exponents although the 
%underling behaviour of the physical quantities is of molecular field type. 

\subsection{Finite size scaling for the energy picture}
The internal energy of a system is given by  
the expectation value (\ref{eq:int_ener}).  
The expression $u_L=u_L(T)$ can be inverted in order to obtain the temperature as a function of the internal 
energy of the system. All physical quantities can then be expressed as functions of the 
internal energy. In the following subsections this transformation 
from the temperature picture to the 
energy picture is discussed for systems displaying either an algebraically or a logarithmically 
diverging specific heat in the thermodynamic limit.

\subsubsection{Algebraically diverging specific heat}
Consider the spontaneous magnetisation near the 
pseudo-critical point of the finite system:
\begin{equation}
  \label{eq:mag_temp}
  m_{sp,L}(t') = A_{\phi}L^{\frac{\beta'-\beta}{\nu}}\sqrt{-t'}
  (1 + B_Lt' + \ldots)
\end{equation}
where a correction term to the 
leading behaviour with the $L$-dependent amplitude $B_L$ is explicitly taken into account. 
From finite size scaling theory one obtains a similar expression for the 
(intensive) zero-field specific heat near $t'=0$ for negative $t'$:
\begin{equation}
  \label{eq:spec_heat}
  c_{h=0,L}(t')= aL^{\frac{\alpha - \alpha'}{\nu}}(1 + 2b_Lt' + \ldots)
\end{equation}
with an $L$-dependent amplitude of the linear correction 
term. Here it is assumed that the singularity of the specific heat of the 
infinite system is characterised by $\alpha \neq 0$. The case of a logarithmic 
divergence is deferred to the next subsection. 
Note also that the specific heat in molecular field approximation does not diverge.
It exhibits a jump singularity, hence one has  
the exponent $\alpha'=0$. 
At the pseudo-critical point of the finite system, i.e.   
$t'=0$, equation (\ref{eq:spec_heat}) reduces to
\begin{equation}
	\label{eq:fss_spez}
	C_L = c_{h=0,L}(0)= aL^{\frac{\alpha - 	                   
\alpha'}{\nu}} \;.
\end{equation}
This relation can be used to estimate the true critical
exponent $\alpha$ from the maxima of the specific heat of the various
finite systems.

As 
\begin{equation}
  c_{h=0,L} = \frac{du(T)}{dT} = \frac{du}{dt'}\frac{dt'}{dT}
  = \frac{1}{T_{c,L}}\frac{du}{dt'}
\end{equation}
integration of (\ref{eq:spec_heat}) gives 
\begin{equation}
  \varepsilon'(t') = u_L(t') - u_{c,L}
\end{equation}
with $u_{c, L}$ being the constant of integration. This constant is just the pseudo-critical 
energy of the finite system. 
The expression for $\varepsilon'(t')$ can be inverted for small $t'$
to give 
\begin{equation}
  \label{eq:temp_ener}
  t' = \frac{1}{T_{c,L}a}L^{-\alpha/\nu}\varepsilon' 
  - \frac{b_L}{T_{c,L}^2a^2}L^{-2\alpha/\nu}\varepsilon'^2 + \ldots
\end{equation}
in the asymptotic limit $\varepsilon' \to 0$.
Plugging this result in equation (\ref{eq:mag_temp}) one finally obtains
\begin{eqnarray}
  \label{eq:mag_ener}
  m_{sp, L}(\varepsilon') & = & A_{\phi}L^{\frac{\beta'-\beta_{\varepsilon}}{\nu_{\varepsilon}}}
  \sqrt{\frac{-\varepsilon'}{aT_{c,L}}} \nonumber \\
& & \left(1 + L^{-\alpha/\nu}\frac{1}{aT_{c,L}}(B_L-b_L)\varepsilon' + \ldots\right)
\end{eqnarray}
with the renormalised exponents $\beta_{\varepsilon}= \frac{\beta}{1-\alpha}$ and 
$\nu_{\varepsilon} = \frac{\nu}{1-\alpha}$. These renormalised critical exponents describe the 
singular behaviour of the various thermostatic quantities of the infinite system in the 
energy picture \cite{Promberger95}. %To see this one solves $c \sim t^{-\alpha}$ for $t$ 
The reformulation also yields the 
finite size scaling relation   
\begin{equation}
   \label{eq:fss}
        L^{\beta_{\varepsilon}/\nu_{\varepsilon}}m_{sp, L}(\varepsilon') = \tilde{\phi}
        \left( \frac{\varepsilon'L^{1/\nu_{\varepsilon}}}{T_{c, L}} \right)
\end{equation}
for the energy picture with a scaling variable that contains the transition temperature of the finite system.
In a completely analogous way the reformulation can be done for other physical quantities like the susceptibility.  
For large systems the pseudo-critical temperature $T_{c,L}$ is close to the critical temperature of the infinite 
system so that is can be replaced by $T_c$. However, as we are dealing with very small clusters the additional 
$L$-dependence in the scaling variable due to $T_{c, L}$ can not be neglected.        
Similar to equation (\ref{eq:fracshift}) one obtains the shift relation for the pseudo-critical 
energy:
\begin{equation}
  \label{eq:fracshift_en}
  e_L = \varepsilon_{c, \infty} -\varepsilon_{c, L} \sim -D_{\varepsilon}L^{-1/\nu_{\varepsilon}} \;.
\end{equation}
The proposed scaling variable for the energy picture and the corresponding scaling relations are tested in Section IV.

In equation (\ref{eq:mag_ener}) the amplitude of the linear correction term contains the amplitudes 
$B_L$ and $b_L$ of the temperature picture. Moreover a prefactor $L^{-\alpha/\nu}$ 
appears which will lead to a suppression of the leading correction term 
compared to the corresponding 
expression (\ref{eq:mag_temp}) in the temperature 
picture if the critical exponent $\alpha$ is positive. This 
suppression is determined by the prefactor together with the 
$L$-dependence of the amplitudes $B_L$ and $b_L$. If the amplitudes grow strongly enough with the 
system size the suppression becomes weaker for increasing $L$.     
Note however that a reduction of the linear corrections in the energy
picture will eventually always take place even if this reduction becomes weaker for
larger cluster sizes.
For a system with a negative exponent the 
correction term in the energy picture is blown up relative to the temperature picture. 
The investigation of this interesting consequence in the $3d$ Heisenberg model as 
an example of a system with $\alpha < 0$ is left to future studies.  

\subsubsection{Logarithmically diverging specific heat}
With slight modifications the above discussed transfer to the energy picture is carried out 
for the case of a logarithmically diverging specific heat $c = - A\ln t$ of the infinite 
system. A prominent example of such a system is the two-dimensional Ising model. 
In this case the amplitude $C_{L}$ (i.e. the maximum) of the finite system 
specific heat varies as \cite{Ferdinand69}	
\begin{equation}
  \label{eq:ampl_spez_log}
  C_{L} = A\ln L^{1/\nu} + \xi = A^{\prime}\ln L + \xi \;.
\end{equation}

\begin{figure}[t]
\centerline{\psfig{figure=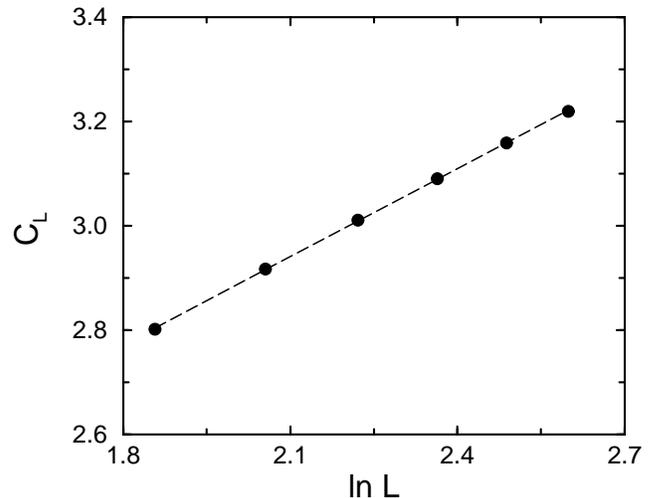,width=3.3in}}
\caption{Specific heat maxima as function of $\ln L$ for small two-dimensional Ising clusters
defined on a square lattice. The linear extension $L$ is given by $L = \sqrt{N}$ where
$N$ is the total number of spins forming the cluster. The dashed line results from
fitting these data to Eq.\ (\ref{eq:ampl_spez_log}), with $A'=0.562$ and $\xi=1.76$.}
\end{figure}

Fig. 5 shows equation (\ref{eq:ampl_spez_log}) for the 
$2d$ Ising model in the extended molecular field approximation. 
The depicted linear behaviour is characterised by the 
slope $A^{\prime} = 0.562$ and the intercept $\xi = 1.76$ with the vertical axis.   
For sufficiently large system sizes the $L$-independent constant $\xi$ can be neglected relative to the 
increasing contribution from the logarithmic term. As relatively small 
systems are considered in the molecular field approximation discussed here this constant 
is explicitly taken into consideration. The arguments used to replace the variable $t'$ by the 
internal energy still apply and give rise to the relation 
\begin{eqnarray}
  m_{sp, L}(\varepsilon') & = & A_{\phi}L^{\frac{\beta'-\beta_{\varepsilon}}{\nu_{\varepsilon}}}
  \sqrt{\frac{-\varepsilon'}{(A^{\prime}\ln L+\xi)T_{c,L}}} \nonumber \\
&& \hspace*{-1cm}\left(1 + \frac{1}{(A^{\prime}\ln L+\xi)T_{c,L}} (B_L-b_L) \varepsilon' + \ldots\right) \;.
\end{eqnarray}
The linear correction term is again diminished by an additional $L$-dependent prefactor as 
already observed for systems with $\alpha > 0$. This form of the magnetisation suggests 
the finite size scaling law 
\begin{equation}
  \label{eq:fss_log}
        L^{\beta_{\varepsilon}/\nu_{\varepsilon}}m_{sp, L}(\varepsilon') = \tilde{\phi}
        \left(\frac{\varepsilon'L^{1/\nu_{\varepsilon}}}{(A^{\prime}\ln L+\xi)T_{c,L}}\right)
        = \tilde{\phi} \left(  \frac{\varepsilon'L^{1/\nu_{\varepsilon}}}
        {C_L T_{c,L}} \right)
\end{equation}
with a slightly modified scaling variable. This modified scaling variable in the energy picture 
is investigated for the two-dimensional Ising model in the next Section. Note however that the factor causing the suppression of the linear 
correction term of the magnetisation is now
integrated into the scaling variable. This has the consequence that the 
reduction of the corrections to scaling in (\ref{eq:fss_log}) are solely controlled by the relative size of the 
amplitudes $B_L$ and $b_L$. Therefore it has to be expected that the reduction of the corrections 
are weaker than those in the case of an algebraically diverging specific heat.

\section{Critical quantities}
In this Section we extract critical quantities like critical exponents or transition 
temperatures of the infinite system from the molecular field data of the corresponding 
finite systems. Apart from the molecular field finite size scaling relations presented in the 
previous Section the notion of the effective critical exponent provides another way to obtain 
critical exponents from finite system quantities \cite{Pleim98}. 

The effective critical exponent $\beta_{eff, N}$ of the order parameter is defined by expressing the 
temperature dependent spontaneous magnetisation $m_N$ in the form 
\begin{equation}
	\label{eq:eff_exp}
	m_N = A\,t^{\beta_{eff, N}(T)} \;,
\end{equation}
 where $t = (T_c-T)/T_c$ and $T_c$ is again the critical temperature of the infinite 
system. The temperature dependent effective exponent is obtained by differentiating relation 
(\ref{eq:eff_exp}): 
\begin{equation}
	\beta_{eff, N}(T) = \frac{d \ln(m_N)}{d \ln t}\;.
\end{equation}
Fig.\ 6 shows the effective exponent of the two-dimensional Ising model for the 
different clusters on the square lattice together with the effective critical 
exponent $\beta_{eff, \infty}$ of the infinite system \cite{Coy}. As the pseudo-critical temperatures of the 
finite systems are above the critical temperature $T_c$ of the infinite system the spontaneous 
magnetisation $m_N(T)$ is finite at $T_c$ and therefore the effective exponent has to vanish at $T_c$. 
It is also observed that the effective exponents of the largest clusters are closer to the corresponding 
curve of the infinite system than those of the small clusters. From the effective exponents estimates 
for the true critical exponent $\beta$ may be extracted by means of linear extrapolation. 
To this end an extrapolated exponent $\beta_{ext, N}$ is defined by 
\begin{equation}
	\label{eq:ext_exp}
	\beta_{ext, N} (T) = \beta_{eff, N}(T) + \frac{d\beta_{eff, N}(T)}{dT} (T_c-T) \;.
\end{equation}
This extrapolated exponent is displayed in Fig.\ 7 for the square lattice again together with 
$\beta_{ext, \infty}$ of the infinite system. 
Note that the effective exponent depends on the critical temperature of the infinite lattice. For models 
with an unknown value of $T_c$ a reliable estimate has 
to be obtained from the finite system pseudo-critical 
temperatures. 
This can be achieved by using the finite size scaling relation (\ref{eq:fracshift}).    

\begin{figure}
\centerline{\psfig{figure=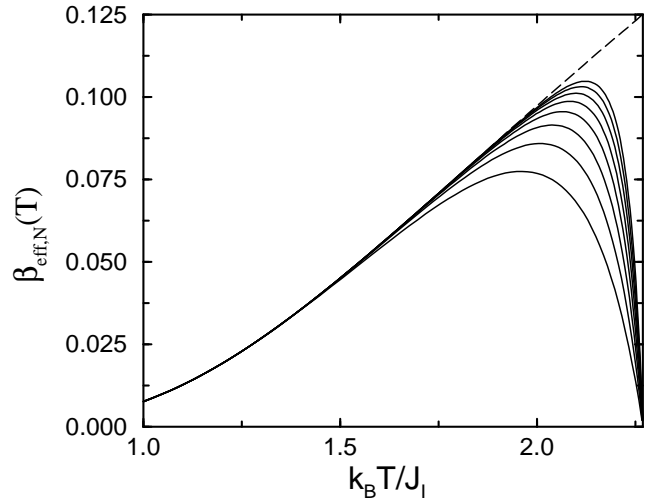,width=3.3in}}
\caption{Effective critical exponents $\beta_{eff,N}$ versus temperature for different
square lattice Ising clusters. The system sizes are the same as those in Fig.\ 2, the lowest
line resulting from the cluster with 13 spins. These curves directly follow from the
logarithmic derivative of the spontaneous magnetisations displayed in Fig.\ 2 with
respect to the temperature. The dashed line is the effective critical exponent
of the infinite system, yielding the critical exponent $\beta = 1/8$ in the limit
$T \longrightarrow T_c$.}
\end{figure}

\begin{figure}   
\centerline{\psfig{figure=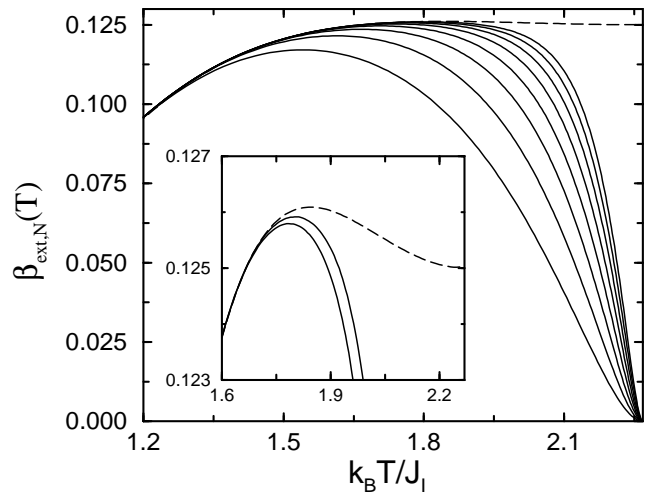,width=3.3in}}
\caption{Temperature dependent extrapolated exponents $\beta_{ext,N}(T)$ derived from the
effective critical exponents shown in Fig.\ 6, see Eq.\ (\ref{eq:ext_exp}). The
dashed line is again the corresponding quantity of the infinite system.
The inset demonstrates a peculiarity of the $2d$ Ising model. The extrapolated
exponent of the infinite system overshoots the asymptotic value 1/8 and this is also true
for the cluster MFA results for $N=145$ and 181.
%As shown
%in the inset this extrapolated exponent presents a slight overshooting above the
%asymptotic value 1/8.
}
\end{figure}

In the next Subsections we investigate our molecular field data for the various systems 
using both finite size scaling relations and effective exponents. 

\subsection{Two-dimensional Ising models}
Using the cluster molecular field approximations we studied two-dimensional nearest
neighbour Ising models
defined on the square and on the triangular lattices as well as an Ising model with 
both ferromagnetic nearest and next-nearest neighbour interactions defined on the square
lattice. The extrapolated exponent from Eq.\ (\ref{eq:ext_exp}) allows us to obtain reliable values
for the order parameter critical exponent $\beta$, as shown in Fig.\ 7 for the square lattice
model with only nearest neighbour interactions.
The best estimate for
$\beta$ is obtained from the value of the temperature where 
the extrapolated exponents for the two largest systems ($N = 145$ and $N = 181$) still
coincide, yielding $\beta \approx 0.126$ which should be compared to the exact
value $\beta_{ex} = 1/8$. This overestimation of only $0.5 \%$ is rather
remarkable considering the smallness of the systems under consideration. Note 
however that the slight overshooting above the true value 1/8 is not an artifact 
of our approximation, it is due to the peculiar behaviour of the exact solution
of the infinite system, see the inset of Fig.\ 7. 

A similar behaviour is also observed for the other two-dimensional Ising models.
In these cases the overshooting is somewhat larger,
yielding a slightly increased value for $\beta$. Thus we obtain the value
$\beta \approx 0.128$ for the triangular lattice, still in very good agreement
with the exact result.

Other critical exponents, as for example the critical exponent $\gamma$
of the susceptibility, might be derived in the same way. In the present study, however,
we restrict ourselves to the small system sizes where the density of states can be computed
exactly. For these systems the extrapolated exponents obtained from the
susceptibility do not yet present a plateau, making an estimation of $\gamma$ rather
tedious. Here, the molecular field finite size scaling theory of Section III is
an interesting alternative for extracting critical quantities, as discussed in
the following. 
 
Let us start by probing the finite size scaling relations (\ref{eq:scaling_mag}) 
and (\ref{eq:fss_log}) of the spontaneous magnetisation in the temperature
as well as in the energy picture. Note that relation (\ref{eq:fss_log}) has to be
used instead of (\ref{eq:fss}) as the specific heat of the two-dimensional model
presents a logarithmic divergence with $\alpha = 0$. 
Plotting the scaled magnetisation as a function of the scaling variable
and setting $L = N^{1/d}$, $d$ being the number of space dimensions,
the best data collapse as derived from a least-square
fit procedure or from a more sophisticated approach \cite{Sen} yields the values
of the critical exponents. The resulting data collapses are shown in Fig.\ 8
for the nearest neighbour Ising model defined on the square lattice where the
smallest systems with 5 and 13 spins have been discarded. Hereby the
critical exponents in the temperature picture take the values $\beta = 0.127$ and
$\nu=1.01$ (see Fig.\ 8a), while in the energy picture (see Fig.\ 8b) the
best data collapse is achieved for $\beta_\varepsilon=0.124$ and $\nu_\varepsilon=1.02$.
This convincingly shows that the finite size scaling relations derived in Section III
are indeed fulfilled by our scaled data.
Again, one has to stress the remarkable agreement between our estimates of the order parameter
critical exponents and the exact value 1/8, based on data obtained for very small system sizes.
%The values of $\nu$ and $\nu_\varepsilon$ differ more strongly from the exact value
%$\nu_{ex} = 1$ of the infinite system. This larger discrepancy 
%can be related to the problematic definition of the linear extension $L$.

\begin{figure}
\centerline{\psfig{figure=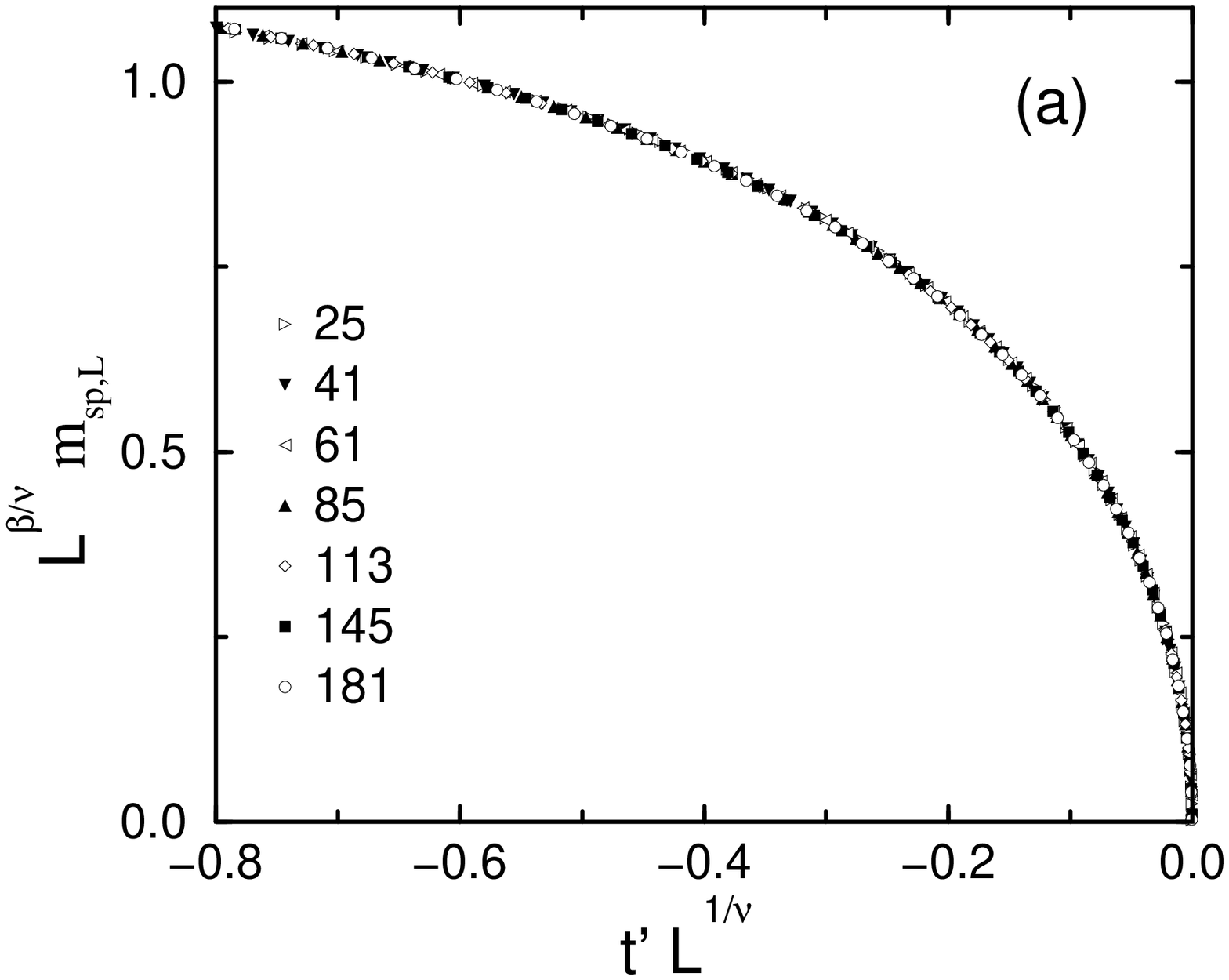,width=3.3in}}
\centerline{\psfig{figure=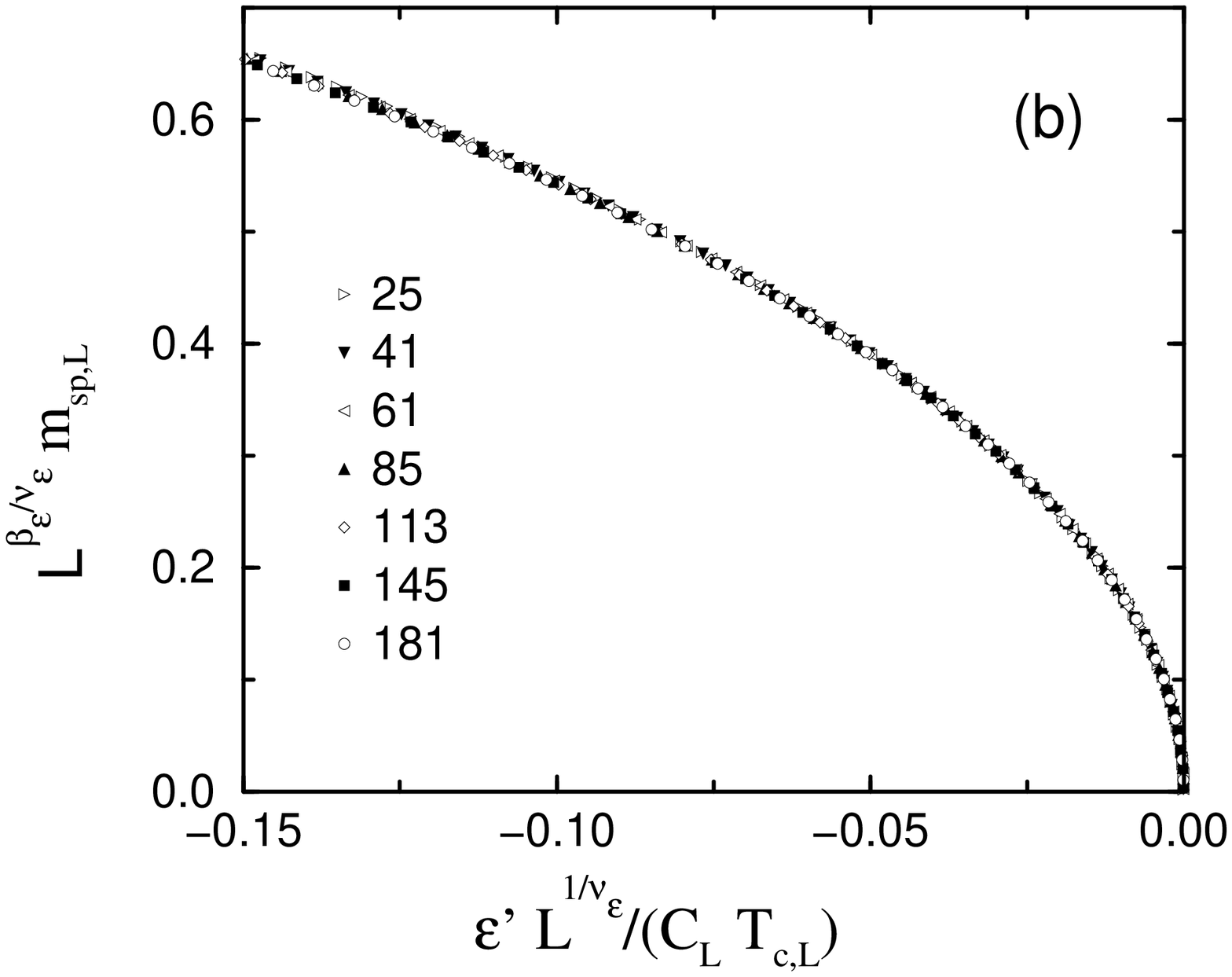,width=3.3in}}
\caption{Scaled magnetisations of Ising clusters defined on a square
lattice, see Fig.\ 2, as function of different scaling variables:
(a) in the temperature picture, see Eq.\ (\ref{eq:scaling_mag}), and (b) in the
energy picture, see Eq.\ (\ref{eq:fss_log}). The best data collapse yields the
following values for the critical exponents: (a) $\beta = 0.127$, $\nu = 1.01$,
(b) $\beta_\varepsilon= 0.124$, $\nu_\varepsilon = 1.02$.}
\end{figure}

Fig.\ 9 displays the scaled susceptibilities
as function of the scaled reduced temperature. The values of the
critical exponents
$\gamma=1.77$ (to be compared with the exact value $\gamma_{ex}=1.75$)
and $\nu=1.01$ again result from the best collapse of the scaled data.
The same can be done in the energy picture where we obtain $\gamma_\varepsilon=1.80$
and $\nu_\varepsilon=1.02$, in reasonable agreement with the exact values.
This shows that the finite size scaling theory enables us to obtain reliable values
for critical exponents also in cases where an analysis of effective exponents does
not succeed.

\begin{figure}
\centerline{\psfig{figure=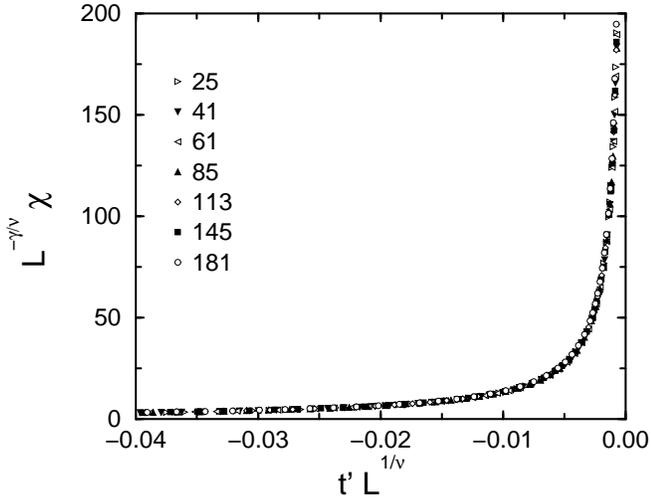,width=3.3in}}
\caption{Scaled susceptibilities obtained for square lattice Ising clusters
in the cluster molecular field approximation, see Fig.\ 4, as function of the scaled reduced
temperature. The values $\gamma = 1.77$ and $\nu = 1.01$ of the critical exponents
result from the best data collapse.}
\end{figure}

At this stage we have to emphasise again that no non-universal quantities 
of the infinite system enter
the molecular field finite size scaling theory. Especially, the exact location of the
critical point is not needed. On the other hand, using the value of $\nu$ obtained from finite size
scaling together with the pseudo-critical temperatures of the three
largest systems listed in Table I, a remarkable good
estimation of the critical temperature follows from Eq.\ (\ref{eq:fracshift}): $k_BT_c/J_I= 2.266$
in excellent agreement with the exact value $k_BT_{c}/J_I= 2/\ln(1+ \sqrt{2}) \approx 2.269$.

Before closing this subsection, we note that almost identical estimates are obtained for
critical quantities when studying the triangular model: $\beta = 0.127$, $\beta_\varepsilon= 0.125$,
$\gamma=1.78$, $\gamma_\varepsilon=1.80$,
$\nu=1.03$, $\nu_\varepsilon=1.03$, and $k_BT_c/J_I= 3.631$ (exact: $3.634$).
The discrepancies are somewhat larger for the square lattice
clusters where the next-nearest neighbour interactions have also been taken into account.
We obtain $\beta = 0.120$ and $\gamma=1.86$, for example, analysing systems with 
up to 81 spins.

\subsection{Three-dimensional Ising model}
When trying to extract critical quantities for the three-dimensional Ising model
from the cluster molecular field approximation, we are faced with two difficulties:
(1) only systems with a very small number of spins ($\leq 57$) can be handled exactly
and (2) the various clusters have different geometries, as explained in Section II.
Especially the second point makes a finite size scaling analysis much more problematic, as discussed
below. But let us start again by looking at the effective exponent derived from the spontaneous
magnetisation. Fig.\ 10 shows the extrapolated exponent obtained for the different system sizes.
The estimated value of 0.316 for the critical exponent $\beta$ is slightly smaller than the 
expected value $0.326$. This is due to the finite-size effects which 
for our small systems set in already at rather low temperatures. 
Fig.\ 10 also displays an oddity due to the different cluster geometries:
on the dashed line the extrapolated exponents of two different system sizes,
$N=27$ and $N=33$, are almost indistinguishable. Looking at the raw data, 
one observes that, by coincidence, due to the difference in the geometries, the 
cluster molecular field approximation yields virtually identical
spontaneous magnetisations for these two clusters.

\begin{figure}
\centerline{\psfig{figure=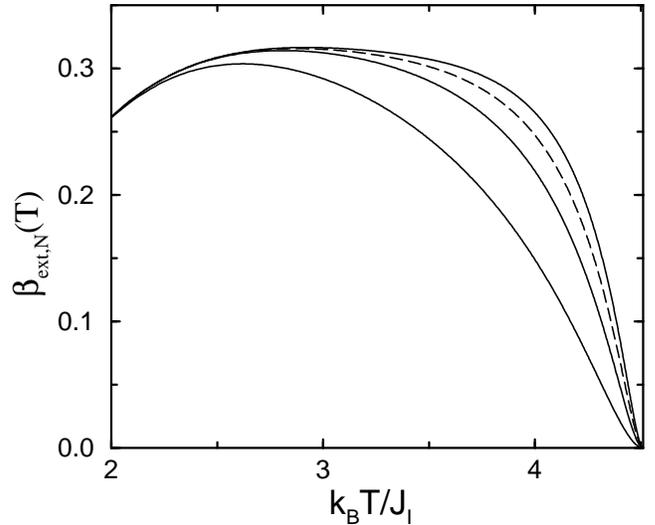,width=3.3in}}
\caption{Extrapolated exponents $\beta_{ext,N}$ versus temperature for different
three-dimensional Ising clusters defined on a cubic lattice. The quantities derived
for systems with 27 and 33 spins are indistinguishable (indicated by the
single dashed line).}
\end{figure}

The difficulties originating from the difference in geometry are
enhanced when looking at the finite size scaling behaviour in three dimensions. 
In fact, the rather naive
identification $L = N^{1/d}$, which works very well in two dimensions 
with identical geometries, is inadequate
when different cluster geometries are involved. Sticking to this definition of $L$,
we are only able to obtain simultaneously a good data collapse for all the
quantities of interest (magnetisation and susceptibility as function of both
reduced temperature and reduced energy) when the sizes $N=27$ and $N=33$ are omitted.
This procedure then yields the following values for the critical exponents:
$\beta = 0.32$, $\beta_\varepsilon=0.41$, $\gamma=1.23$, $\gamma_\varepsilon=1.45$,
$\nu=0.63$ and $\nu_\varepsilon=0.69$. Whereas in the temperature picture the
values of the critical exponents agree nicely with the values given in the
literature ($\beta = 0.326$, $\gamma=1.24$, $\nu=0.63$), some discrepancies
show up in the energy picture (for example, one expects 0.367 for $\beta_\varepsilon$),
even if the overall agreement is still very satisfactory with regard to the smallness
of our systems.

A possible way to circumvent some of the difficulties arising from the difference
in cluster geometry would be to eliminate $L$ altogether in Eq.\ (\ref{eq:scaling_mag})
resp.\ (\ref{eq:fss}) by exploiting \cite{Suz86b} the relation (\ref{eq:fracshift})
resp.\ (\ref{eq:fracshift_en}). However, this would introduce the necessity to determine
precisely the location of the critical point beforehand.

An estimate of the critical exponent $\alpha$ of the infinite
three-dimensional Ising system may be obtained by applying relation (\ref{eq:fss_spez}) for
the maxima of the specific heat of the finite cluster systems. Using the
system sizes 19 and 57 one obtains the value $\alpha = 0.132$, again in reasonable 
agreement with the literature value $\alpha = 0.109$.

In Section III we already discussed briefly the possible reduction of the correction to scaling 
terms when going from the temperature to the energy picture. 
Denoting the amplitude of the linear correction term in  (\ref{eq:mag_ener}) 
by 
\begin{equation}	
	K_L = L^{-\alpha/\nu}\frac{1}{a}(B_L-b_L)
\end{equation}
the ratio $B_L/K_L$ may serve as a measure of the suppression of the linear correction term 
in (\ref{eq:mag_ener}) compared to the corresponding term in (\ref{eq:mag_temp}). Note that the 
pseudo-critical temperature is not integrated into this ratio as $T_{c,L}$ appears in the scaling variable 
of the energy picture. 
For the magnetisation of the  
three-dimensional clusters this ratio is of the order 8. Similarly the ratio $B_L/(B_L-b_L)$ describes the 
reduction of the correction term for a system with a logarithmically diverging specific heat. Again the 
factors appearing in the scaling variable of (\ref{eq:fss_log}) are excluded from the ratio. For the 
two-dimensional Ising system in the cluster molecular field approximation this ratio is of the order 6. 
In both cases the corrections to scaling are suppressed in the energy picture as expected from the 
considerations in Section III. 
It is also observed in our data that the suppression decreases slightly with increasing system size 
for the range of extensions $L$ investigated in our cluster molecular field approximation. 
    
Finally, we again get a very good estimate of the critical temperature 
from Eq.\ (\ref{eq:fracshift}). Inserting the value $\nu=0.63$ coming from the
best data collapse of the scaled data as well as the pseudo-critical temperatures
of the systems with 7, 19, and 57 spins, we find $k_BT_c/J_I=4.5208$ which
should be compared to the literature value $k_BT_c/J_I=4.5115$.

\subsection{Three-state Potts models}
Let us close this Section by discussing three-state Potts models on square 
and triangular lattices. In these cases the cluster molecular field approximation
displays a discontinuous transition for all finite clusters, as
illustrated in Fig.\ 3 by the jump of the order parameters. At first sight
this fact seems to rule out the possible determination of critical quantities
from our small systems. However, we show in the following that our methods
of analysing the data in principle still allow us to obtain reliable values for the
critical quantities.

\begin{figure}
\centerline{\psfig{figure=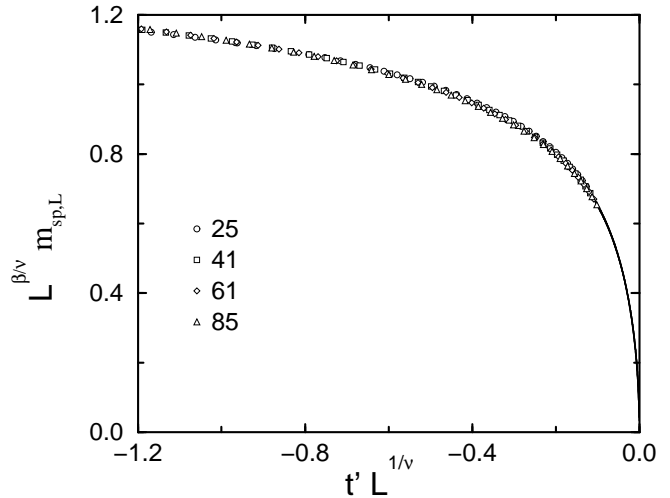,width=3.3in}}
\caption{Scaled magnetisations of Potts clusters defined on a square
lattice, see Fig.\ 3, as function of the reduced temperature scaling variable.
Real data are shown by symbols, data obtained from a prolongation of the polynomial fits 
to $m_{sp}=0$, see main text, by the full lines. The best data collapse is achieved for the
critical exponent values $\beta=0.11$ and $\nu=0.76$.}
\end{figure}

In the present situation the analysis of effective critical exponents has the
great advantage that only data computed for temperatures lower than the 
critical temperature of the infinite system are involved. Therefore the
discontinuous behaviour of our small systems 
observed at temperatures above the critical temperature
does in this case not interfere with the determination of the
critical exponents. Thus the extrapolated critical exponents computed
for square lattice systems with up to 85 spins yield the value $\beta = 0.104$
for the order parameter critical exponent, in good agreement with the 
exact value $\beta_{ex} = 1/9$. On the other hand the molecular field finite size
scaling analysis is deeply affected by the discontinuity of the phase transitions as here
data obtained in close vicinity to the transition points play a prominent role.
In fact, the main problem lies in the values of the pseudo-critical temperatures
which enter the definition of the scaling variables. Using the transition temperatures
given by the kinks in the cluster molecular field free energies, 
see inset of Fig.\ 3, does not lead to a
collapse of the scaled data. A better result is obtained by fitting the spontaneous
magnetisations of the finite systems to polynomials of second order and then inserting
the temperatures where these polynomial functions
hit the temperature axis into the scaling variables. This approach leads to an 
acceptable data collapse for the
order parameter, as shown in Fig.\ 11, the critical
exponents thereby taking the values $\beta=0.11$ and $\nu=0.76$. The value
of $\beta$ again nicely agrees with the exact result but a slight discrepancy prevails
for $\nu$ as the exact value is 5/6. With regard to the crudeness of our
approach, this is still a very satisfactory result. 
However, we did not succeed in extracting the critical exponent $\gamma$ from the
susceptibility data in a similar way.
Finally, inserting the
value $\nu=0.76$ into Eq.\ (\ref{eq:fracshift}) together with the temperatures obtained
from the polynomial fits, we end up with the estimate $k_BT_c/J_P=0.998$
for the critical temperature, very close to the exact value $k_BT_c/J_P 
= 1/\ln(\sqrt{3} + 1) \approx 0.995$.

\section{Conclusion}
The advantages and shortcomings of molecular field theory are well known. On the
one hand this approximation provides a rather simple way of obtaining analytical
results even in complicated systems. Usually, the results of the molecular field
approximation are qualitatively correct. On the other hand, however,
large quantitative discrepancies are found when comparing these predictions
with exact results or with estimations obtained by numerical methods and/or 
more sophisticated analytical approaches. This is especially the case when
considering systems with a continuous phase transition as the molecular field
approximation always yields classical values for the critical exponents 
irrespective of, for example, the dimensionality of the system. Furthermore,
critical temperatures are usually overestimated by a considerable amount.

In this work we have shown that the correct values of critical quantities
can nevertheless be obtained from systematic improvements of the molecular
field approximation. We thereby insert spin clusters of variable sizes into
a homogeneously magnetised background and determine the molecular field
acting exclusively on the border spins by imposing a condition of homogeneity
in the central part of the cluster. The interactions inside the clusters are
treated exactly by computing the density of states as function of the internal
energy and of the interaction energy with the molecular field.

The data computed in this cluster molecular field approximation
can then be analysed in various ways in order to extract the values 
of critical quantities. We have discussed in this work two rather 
complementary approaches, one approach involving effective critical exponents where only
data located at temperatures below the critical temperature of the infinite
system enter, the other being a molecular field finite size scaling theory (which
in the temperature picture is related to that developed by Suzuki 
\cite{Suz86a,Suz86b,Suz86c,Suz87,Kat87,Hu87}) where the
true critical exponents result from the best collapse of the rescaled data.
In the latter approach the data located in close vicinity to the pseudo-critical
temperatures are of main interest. 

With these methods data resulting from the cluster molecular field treatment
of various two- and three-dimensional Ising models as well as of two-dimensional
three-state Potts models have been investigated. In all these cases we have shown
that critical temperatures and critical exponents can be determined with a surprisingly
high precision with regard to the smallness of our spin clusters. In two dimensions
we have the nice situation that clusters of various sizes with the same shape can
be handled by the available computational capacities. This is different in three
dimensions where our computational resources do not allow us to handle spin
clusters of various sizes having the same geometry. In fact, these restrictions 
are imposed by our choice to compute the density of states in a numerically exact manner.
In principle larger system sizes could be investigated by Monte Carlo simulations, using one
of the highly efficient methods for computing the density of states which have been
developed recently \cite{Hueller02}. However, very good numerical data
would be needed in order to obtain the strength of the molecular field as a 
function of the temperature from Eq.\ (\ref{eq:selbstkonsist}). Especially the accurate  
determination of the pseudo-critical temperature requires a density of states of extremely high
precision not easily achievable in a computer simulation.
We therefore expect that data obtained from Monte Carlo simulations may 
not permit a molecular field finite size scaling analysis. However, they
should be very useful for the investigations of effective critical exponents,
especially when dealing with quantities like the susceptibility or the
specific heat for which the small systems considered in this work did not allow
a determination of the values of the critical exponents.

Further progress along the lines developed here can be achieved by noting that
the homogeneity condition in the central part of the cluster renders our cluster
molecular field approximation superior to more traditional approaches. However,
the spin expectation value is still not identical in the whole cluster. Here
a scheme developed by Galam \cite{Gal96} could be very useful. In that cluster 
molecular field scheme a cluster 
consisting of the central spin and
of its $z$ neighbours is constructed such that the homogeneity is extended
to the whole cluster. It is, however, not immediately obvious how to generalise
this scheme to larger clusters containing not only the central spin and
its nearest neighbours. Nevertheless we expect that the construction of
homogeneously magnetised clusters (be it by the Galam scheme or by a different one)
would further improve the systematic cluster molecular field approximation 
and yield even better values for the critical quantities.

\section*{Appendix }
In this appendix we describe how the density of states is evaluated in a 
numerically exact way. 
For simplicity consider a square lattice which consists of $L$ rows of 
length $L$. In our case the density of states depends on two extensive quantities, the 
interaction energy $E$ of the spins inside the cluster and the interaction 
energy $B$ of border spins with spins $\sigma_k = 1$ outside the cluster. In other physical contexts the density of 
states can depend on other extensive quantities, e.g. the interaction energy and the magnetisation. The 
density of states $\Omega(E, B)$ of the cluster is worked out by successively 
performing the trace over the $q^L$ states of a row of spins. A similar method 
has already been discussed in the literature for a density of states depending only 
on one extensive parameter \cite{Bin,Cre,Kim}. Let $k = 1, \ldots, q^L$ denote the configurations 
of the row of spins and $l = 1, \ldots,L$ the rows of the cluster. Suppose 
$l$ rows of the cluster are already built up and the density of states of the 
incomplete cluster is given by $\omega_l(k, E, B)$. If the next row is added 
additional internal and external interactions of this row with the already built up 
cluster and the spins outside have to be included. Let ${\mathcal{E}}_{k^{\prime}, k}$ 
denote the interaction energy of the microstate $k^{\prime}$ of the last row of the yet incomplete cluster 
with the microstate $k$ of the row to be added in the next step. Similarly ${\mathcal{B}}_{k}$ denotes 
the additional external energy. Of course ${\mathcal{B}}_{k}$ does not contain the 
contributions of those bonds of the new row which will become internal bonds when the 
next row is added. Then the density of states of the extended 
cluster consisting now of $l+1$ rows is given by 
\begin{eqnarray}
  \omega_{l+1}(k, E, B) & = &\sum_{k^{\prime}, E^{\prime}, B^{\prime}}
     \omega_{l} (k^{\prime}, E^{\prime}, B^{\prime}) \nonumber \\
 &&  \delta(E, E^{\prime}+{\mathcal{E}}_{k^{\prime}, k})
     \delta(B, B^{\prime}+{\mathcal{B}}_{k}) \;.
\end{eqnarray}
The density of states of the complete cluster is then obtained as
\begin{equation}
  \Omega(E, B) = \sum_{k} \omega_L(k, E, B) \;.
\end{equation}
The described method is easily generalised to densities of states depending 
on more than two parameters.

%%%%%%%%%%%%%%%%%%%%%%%%%%%%%%%%%%%%%%%%%%%%%%%%%%%%%%%%%%%%%%%%%%%%%%%%%%%%%

\end{document}